\documentclass[12pt]{spieman}  % 12pt font required by SPIE;
\usepackage{amsmath,amsfonts,amssymb}
\usepackage{graphicx}
\usepackage{setspace}
\usepackage{tocloft}
\usepackage{wrapfig}
\usepackage{xcolor}

\newcommand\xlos{\hat{x}_{\mathrm{LOS}}}

\newcommand\zlos{\hat{z}_{\mathrm{LOS}}}
\newcommand\xpld{\hat{x}_{\mathrm{PLD}}}
\newcommand\ypld{\hat{y}_{\mathrm{PLD}}}
\newcommand\zpld{\hat{z}_{\mathrm{PLD}}}

\title{Optimized Observation Sequencing in Low-Earth Orbit with the SPHEREx Survey Planning Software}
\author[a,*]{Sean~Bryan}
\author[b,c]{James~Bock}
\author[b]{Thomas~Burk}
\author[b,c]{Tzu-Ching~Chang}
\author[b,c]{Brendan~P.~Crill}
\author[c]{Ari Cukierman}
\author[b,c]{Olivier~Dor\'e}
\author[b]{C.~Darren~Dowell}
\author[e]{Gregory~Dubois-Felsmann}
\author[b]{Beth~Fabinsky}
\author[b,c]{Sergi~Hildebrandt-Rafels}
\author[c]{Howard Hui}
\author[b]{Kyle~Hughes}
\author[c]{Phillip~Korngut}
\author[a,f]{Philip~Mauskopf}
\author[a]{Julian~Mena}
\author[c]{Chi Nguyen}
\author[e]{Milad~Pourrahmani}
\author[d]{Dustin~Putnam}
\author[b]{Keshav Ramanathan}
\author[b]{Flora Ridenhour}
\author[a]{Cody~Roberson}
\author[b]{Amy~Trangsrud}
\author[b]{Stephen~Unwin}
\author[a]{Pao-Yu~Wang}
\author[$~$]{ and the SPHEREx Team}
\affil[a]{School of Earth and Space Exploration, Arizona State University, 781 Terrace Mall, Tempe, AZ 85287 USA}
\affil[b]{Jet Propulsion Laboratory, California Institute of Technology, 4800 Oak Grove Drive, Pasadena, CA 91109 USA}
\affil[c]{Division of Physics, Mathematics, and Astronomy, California Institute of Technology, 1200 E California Blvd, Pasadena, CA 91125 USA}
\affil[d]{BAE Systems, 1600 Commerce St, Boulder, CO 80301 USA}
\affil[e]{Infrared Processing and Analysis Center, 1200 E California Blvd, Pasadena CA 91125 USA}
\affil[f]{Department of Physics, Arizona State University, 550 E Tyler Drive, Tempe, AZ 85287 USA}

\cftpagenumbersoff{figure}
\cftpagenumbersoff{table} 
\begin{document} 
\maketitle

\begin{abstract}
SPHEREx is a NASA infrared astronomy mission that launched on March 12th, 2025 and is operating successfully in low-Earth orbit (LEO). The mission is currently observing the entire sky in 102 spectral channels in four independent all-sky surveys and also achieves enhanced coverage in two deep fields. This data will resolve key science questions about the early universe, galaxy formation, and the origin of water and biogenic molecules. In this paper, we describe the survey planning software (SPS) that enables SPHEREx to observe efficiently while mitigating a range of operational challenges in LEO. Our optimal target selection algorithm achieves the required high coverage in both the All-Sky and Deep Surveys. The algorithm plans observations to stay within our time-varying allowable pointing zone, interleaves required data downlink passes, and mitigates outages due to the South Atlantic Anomaly and other events. As demonstrated by the sky coverage achieved in the first SPHEREx public data release, our approach is performing well in flight. The SPHEREx SPS is a key new capability that enables the mission to deliver groundbreaking science from LEO.
\end{abstract}

% Include a list of up to six keywords after the abstract
\keywords{satellite operations, astronomy, remote sensing, optimal sensor tasking, infrared, spectroscopy}

% Include email contact information for corresponding author
{\noindent \footnotesize\textbf{*}\linkable{sean.a.bryan@asu.edu} }

%\begin{spacing}{2}   % use double spacing for rest of manuscript, uncomment for journal submission

\section{Introduction}
\label{sect:intro}  % \label{} allows reference to this section
SPHEREx (Spectro-Photometer for the History of the Universe, Epoch of Reionization and Ices Explorer) is a spectral survey mission\cite{bock25,alibay23,crill20} that is conducting all-sky and targeted deep surveys during 25 months of nominal science operations to address key science goals\cite{dore16} in cosmology, astrophysics, and astronomy. The mission began routine science operations in May 2025, and made the first of several public data releases\cite{spherex25} in July 2025. SPHEREx provides the first near-infrared all-sky spectral survey for use by the broader astronomical community. The mission is operating from low-Earth orbit (LEO) in a Sun-synchronous polar orbit. SPHEREx is subject to a number of time-varying and interacting geometric avoidance requirements that form the allowable pointing zone of the observatory. In addition, SPHEREx samples the entire sky in 102 near-infrared bands by stepping the telescope boresight along the spectrally-varying axis of a set of linear variable filters (LVFs) installed over its detectors. To meet the spectral coverage requirements for the mission while simultaneously meeting the avoidance requirements, we developed the Survey Planning Software (SPS) to generate optimal observing plans for operations. The software accepts an orbit predict as well as the sky coverage achieved to date as the inputs, and outputs an observing plan for the coming observing period of half of a week. Our observation planning period is chosen to minimize the impact of small accumulating orbit predict error by letting us use an updated orbit predict for each successive observing period.

\begin{figure}
\begin{center}
\includegraphics[width=0.85\textwidth]{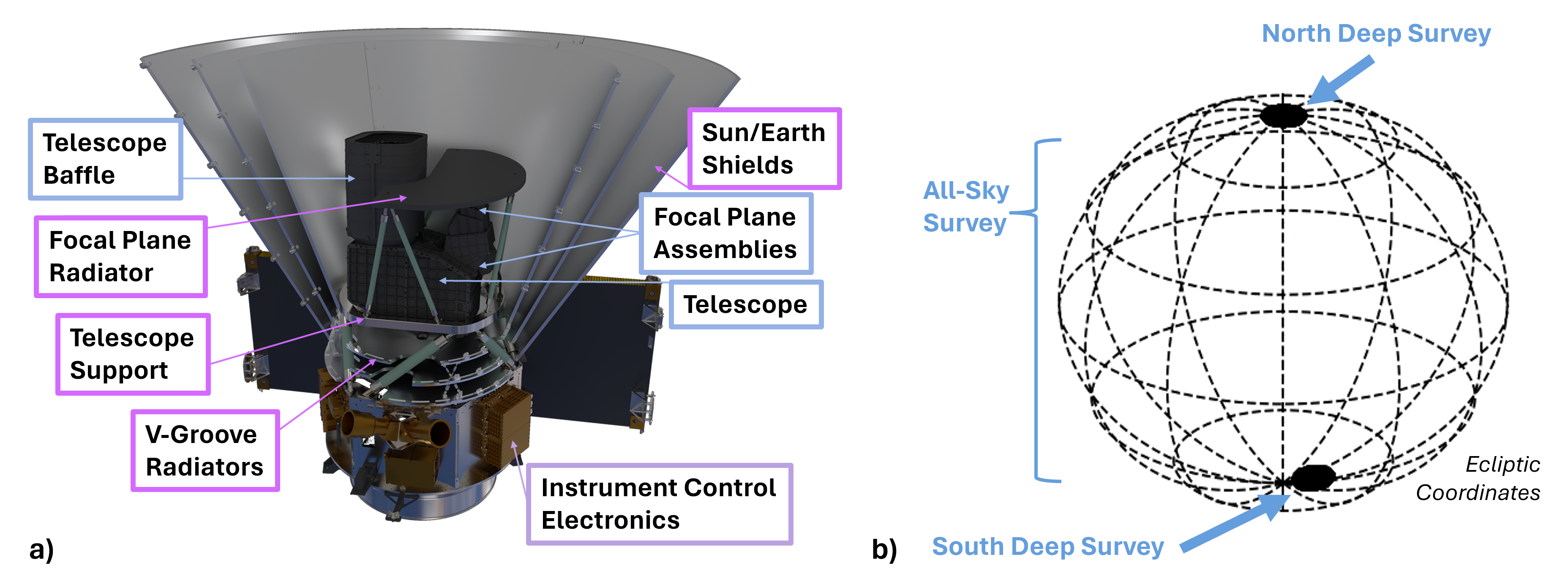} 
\end{center}
\caption 
{ \label{fig:survey_concept}
\textit{a.} Cutaway drawing of the SPHEREx observatory. The telescope (center top) is protected from illumination by the Sun and Earth with the Sun/Earth shield (top). The telescope is tilted Sunward inside the Sun/Earth shield to improve the placement of allowable pointing zone relative to our two deep fields. The spacecraft section (bottom) houses many key components for the mission, including the solar array and instrument control electronics.
\textit{b.} Conceptual overview of the SPHEREx Survey. SPHEREx images the entire sky in 102 spectral channels in the All-Sky Survey, and also delivers improved sensitivity by making repeated observations in two deep fields located near the ecliptic poles. The location of the deep fields is driven by our Sun-synchronous polar orbit which delivers frequent repeated access to these field locations. The southern deep field location is offset from the South Ecliptic Pole (SEP) to avoid overlap with the bright Large Magellanic Cloud as well as certain bright stars.} 
\end{figure}  

Many space missions have needed to address the optimal sensor tasking problem\cite{fabinsky08} but none match the level of complexity of the SPHEREx observing constraints. All-sky surveys such as Planck \cite{planck11} and WMAP \cite{bennett03}, as well as the transformative JWST \cite{gardner06}, decided to incur the additional cost and complexity to fly to the L2 Lagrange point. Since there are fewer constraints on the observatory attitude at L2, Planck and WMAP used repeating geometric observing patterns to meet their observing requirements. Hubble \cite{hst19} operates in LEO which enabled the mission to find a broad range of synergies with the NASA Space Shuttle system for launch and maintenance. The Hubble team developed SPIKE \cite{johnston90} to plan their observations. While this software provides an optimal schedule to observe a pre-selected list of targets, it is not capable of planning an all-sky spectral survey such as SPHEREx. WISE \cite{wright10} conducted all-sky survey operations from LEO using a repeating geometric scan strategy. This approach is robust and delivers all-sky coverage for the 4 spectral bands in the instrument, but unfortunately this approach cannot be adapted for SPHEREx since the WISE flight system features a rotating scan mirror, and simultaneous observation of all 4 of its bands, driving its simple scanning strategy. We leveraged heritage from the WISE survey planning process, but we developed new survey software for SPHEREx since no existing software met the requirements set by the SPHEREx instrument design.

The SPS is written in python and leverages several standard open-source modules\cite{numpy,scipy,healpix,astropy,pandas,spiceypy} to implement our novel optimal sensor tasking algorithm. The SPS plans efficient science observations, pauses science operations for downlink, mitigates outages due to passage through the radiation environment of the South Atlantic Anomaly, and selects a safe place to point if no science targets are observable. The software performed well in integration testing with the rest of the ground system software for the mission before launch, is currently running well using the actual trajectory data we obtained after launch, and successfully planned the observations contained in our first SPHEREx public data release \cite{spherex25}.

In this paper, we present the features and current status of the SPHEREx survey planning software. Several of the core concepts were described in an earlier paper\cite{spangelo15}, and here we describe the current observation strategy, production software, and in-flight performance. We developed the main SPHEREx target list that delivers the required all-sky and deep field coverage. While staying in our allowable pointing zone defined by several time-varying avoidance angle constraints, we use our optimal target selection algorithm to select the order in which each science target should be observed. In mission-long simulations with a nominal orbit predict, our approach yields the required survey coverage with margin.

\section{The SPHEREx Survey}

As illustrated in Figure~\ref{fig:survey_concept}, SPHEREx delivers all-sky coverage and also two deep fields. The All-Sky Survey covers the entire sky in all 102 spectral channels in the instrument, yielding an unprecedented legacy data set for the community that meets a broad range of current and future science needs. The two deep fields yield data that will address the extragalactic background light science goal of our mission, and will enable other future scientific investigations as well.

SPHEREx has a single observing mode, and as described in Section~\ref{sec:optimal}, the SPS observes both the All-Sky and Deep Surveys each orbit to achieve optimal coverage in both. The instrument \cite{crill20} includes two focal plane assemblies, each consisting of three detectors. Each detector images a (3.5 deg)$\times$(3.5 deg) field with 6.15''$\times$6.15'' pixels, and is spaced from neighboring detectors by 0.25 deg such that the field of view imaged with a given exposure is (3.5 deg)$\times$(11 deg). Each detector array is integrated with an LVF to define the passbands. The two focal plane assemblies image simultaneously via a dichroic beam splitter. In data analysis, we divide each individual LVF into 17 horizontal sections, with a given section corresponding to an independent spectral channel with a spatial extent of approximately (11.8 arcminutes)$\times$(3.5 deg) on the sky. The result is (2 focal plane assemblies)$\times$(3 detector-LVF pairs)$\times$(17 channels per LVF) = 102 near-infrared passbands with band centers ranging from 0.75 to 5 microns.

To achieve full spatial and spectral coverage for both the All-Sky and Deep Surveys, the SPS maintains two lists of pointing targets, one for all-sky spectral coverage, and one for frequently repeated spectral coverage in the deep fields.  Our target list puts targets in target groups, with attitudes separated by 11.8 arcminutes corresponding to a single spectral channel. The 11.8-arcminute maneuver between targets within a target group is called a ``small step.'' To move between target groups, SPHEREx executes ``large slew'' maneuvers of up to 70 degrees. As described in Section~\ref{sec:optimal}, the SPS selects the next target group to minimize the number of time-expensive large slews, and maximize the number of observations made within a target group before it moves out of the allowable pointing zone. In this section, we describe the All-Sky and Deep target lists that the SPS uses.

\subsection{All-Sky Target List}
\label{sect:all_sky_tgt_list}
The All-Sky Survey needs to deliver uniform spatial and spectral coverage across the entire sky. To do this efficiently and avoid unnecessary repeat coverage, we developed a main target list with groups of targets equally spaced in solid angle across the entire sky, and with overlap to mitigate pointing error and optical distortion effects.

\begin{figure}
\begin{center}
\includegraphics[width=0.75\textwidth]{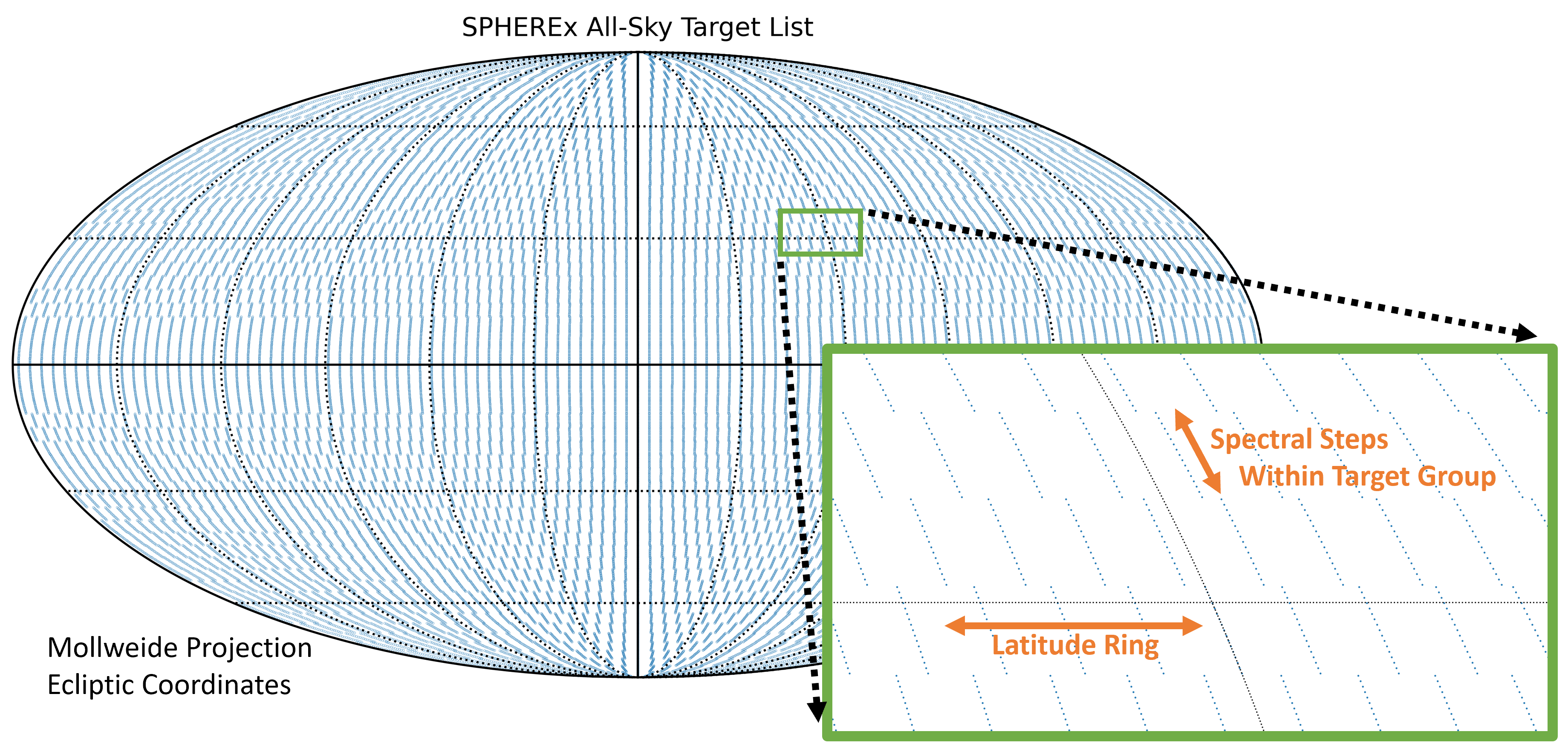} 
\end{center}
\caption 
{ \label{fig:allsky_target}
SPHEREx All-Sky Survey target list. The target list distributes target groups equally in solid angle across the entire sky along latitude rings. Within each target group, there are spectral steps along the spectral axis of the focal plane.} 
\end{figure} 

A single target group consists of 17 pointings at an identical ecliptic longitude and spaced by one channel width in ecliptic latitude. For any single spectral channel, completing a target group yields a square region of full coverage in that single spectral channel. To achieve full coverage in all spectral channels across the entire sky, we distribute these target groups in a great circle around the equator at all longitudes to form a single latitude ring. We construct similar latitude rings at great circles at all latitudes, naturally with fewer target groups in the latitude rings at higher latitude. To mitigate pointing error as well as distortion in the optical system, we overlap the latitude rings, and also overlap the target groups within each latitude ring. This yields an All-Sky target list that achieves full spectral and spatial coverage, efficiently distributes observing time evenly in solid angle across the sky, and contains overlap as mitigation and redundancy.

To efficiently align all observations in the All-Sky Survey, we always observe with the observatory rotated about the telescope boresight such that the wide axis of the field of view aligns with lines of constant ecliptic latitude. This rotation angle is the ecliptic position angle (PA) of the observatory, and is measured relative to the line of constant ecliptic latitude passing through the telescope boresight. To meet the solar panel illumination requirements, during ascending halves of the orbit we align the observatory with a PA of 0$^\circ$, and during descending halves of the orbit the observatory is aligned with a PA of 180$^\circ$. This also plays a key role in the definition of the surveys. Survey 1 is composed of data taken during the first 12.5 months of science operations with PA=0$^\circ$, survey 2 is composed of data taken during the first 12.5 months of science operations with PA=180$^\circ$, survey 3 is composed of data taken during the second 12.5 months of science operations with PA=0$^\circ$, and survey 4 is composed of data taken during the second 12.5 months of science operations with PA=180$^\circ$.

Each target in the All-Sky Survey is spaced by an angle on the sky that corresponds to one spectral channel in the LVF. To enable internal data consistency checks, and to ensure full spectral coverage even of narrow spectral features, Surveys 3 and 4 are shifted on the sky relative to Surveys 1 and 2 by an angle on the sky that corresponds with half of a spectral channel. This shift Nyquist-samples the spectra taken across the entire sky, twice, when data is combined across the full mission.

\subsection{Deep Survey Target List}

SPHEREx conducts a Deep Survey primarily focused on our extragalactic background light science (EBL) investigation. Thanks to the sun-synchronous polar orbit, the polar regions are observable every orbit. This naturally lead us to select two deep fields: one located at the north ecliptic pole (NEP), and the other near the south ecliptic pole (SEP). The southern deep field is offset slightly from the pole to avoid foregrounds and bright stars. As shown in Figure~\ref{fig:deep_coverage}, our target list yields good coverage in both the northern and southern deep fields.

The Deep Survey target list was developed to meet the power spectral sensitivity required by the EBL investigation, discussed in Section~\ref{sec:coverage}. Our strategy is to design a repeatable tiling pattern that, together with sky rotation, achieves a deep-field area and depth that delivers the required power spectral sensitivity. To improve uniformity in mosaics made from our deep field images, we apply a random but known sub-pixel shift to each pointing in the deep fields. We developed both an analytical formalism and simulation software to guide the design and validate the power spectral sensitivity performance. We found multiple tiling patterns that yield similar required sensitivity performances, and subsequently considered secondary figure-of-merits such as deep-field size, depth distribution, tiling cadence, ease of scheduling, and aid with systematics examination. 

Our selected tiling pattern corresponds to the on-sky angular size of one detector array, $3.5^{\circ} \times 3.5^{\circ}$, while sky rotation turns the square pattern into a circular deep-field coverage, with a central high-density coverage a diameter of $\approx 3.5^{\circ} \times \sqrt{2} \simeq 5^{\circ}$ that tapers outwards radially. The total deep-field area that is fully spectrally sampled is about $\approx 100$ deg$^2$ per field. As we define 17 spectral channels per detector length, a fully spectrally sampled tile takes a total of $(17 + 16) \times 3 = 99$ small-step spacecraft movements, which in practice takes several days to complete. This sets the deep-field tile observation cadence that may be of use for additional time-domain science cases, although we note that the deep fields are visited every 98-minute orbit with a few small steps. In addition, each tile corresponds to a slightly offset sky area, rotated about the pole in the NEP case. As discussed, sky rotation helps construct the deep-field sky coverage area and fills in previously observed region that has only partial spectral sampling. The resulting surface brightness sensitivity, if counting only fully spectrally sampled region as SPHEREx defines it, is therefore increased and is deeper than what can be achieved by the sum of small-step integrations.

The pointing targets that make up this pattern at each pole are generated by a separate tool from the core survey software. The pattern is constructed by stepping the boresight along the spectral direction to create target groups similar to those in the All-Sky Survey. Unlike the All-Sky Survey, in the Deep Survey target list these target groups are highly overlapping to yield deep repeated coverage. The SPHEREx science pipeline combines these repeated observations into a mosaic for each spectral band in the instrument, and these mosaics are then processed further to yield the spatial power spectrum at each spectral band. As we discuss in Section~\ref{sec:coverage}, the as-planned Deep Survey yields coverage that meets the required spatial power spectrum sensitivity with margin by the end of science operations.

\section{Operational Constraints}
\label{sec:avoidance}

\begin{figure}
\begin{center}
\includegraphics[width=0.55\textwidth]{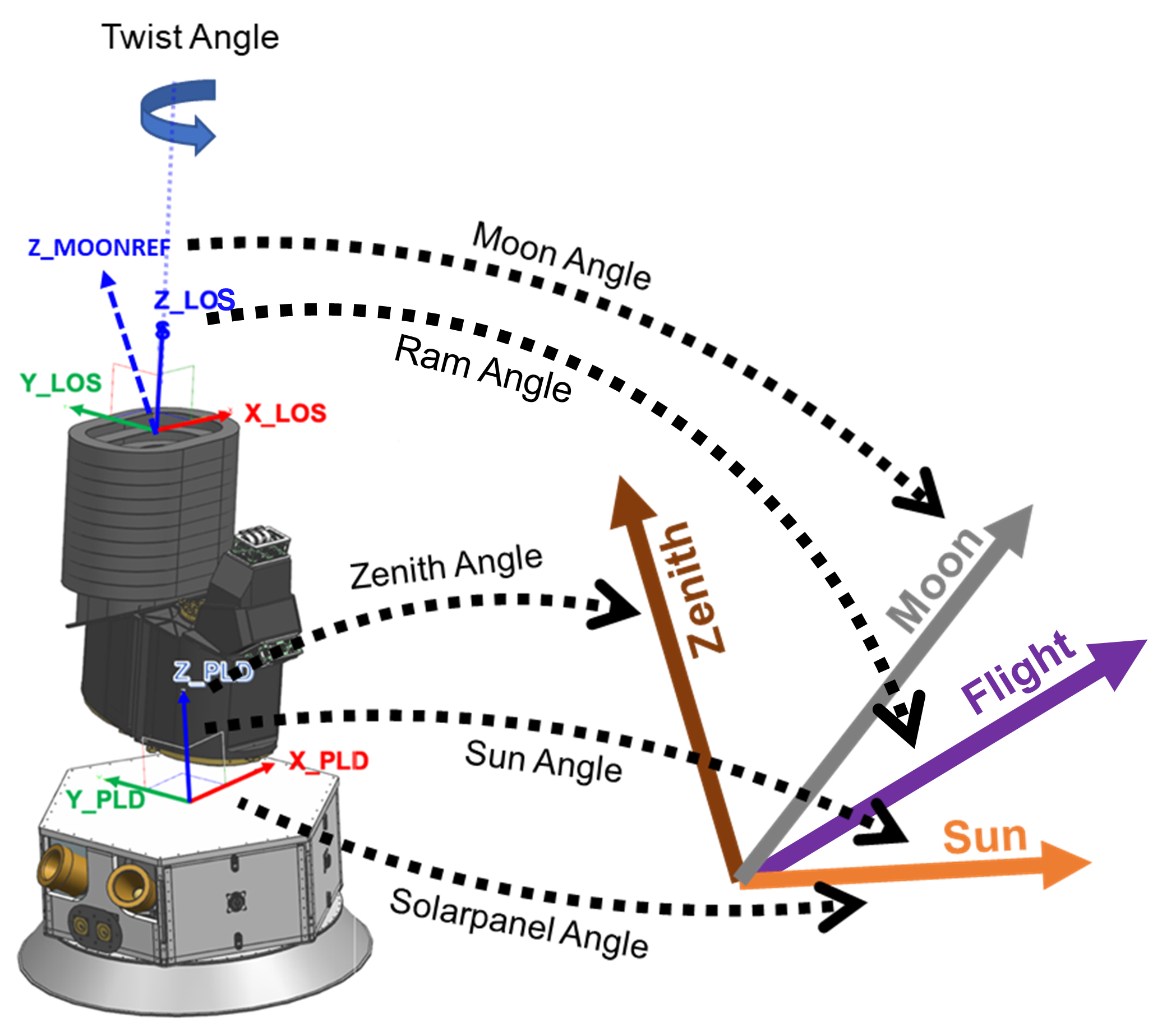} 
\end{center}
\caption 
{ \label{fig:avoidance_angles}
Avoidance angles defining the SPHEREx allowable pointing zone. The payload (PLD) body axes (bottom left) are shown as red, green, and blue arrows over the CAD rendering. The telescope (LOS) body axes (top left) are tilted Sunward within the Sun/Earth shield (not shown here, shown in Figure~\ref{fig:survey_concept}) by 8 degrees to improve the position of the allowable pointing zone relative to our deep fields. As shown (doted black lines), avoidance angles are defined in our requirements from the PLD/LOS body axis vectors to the Earth, Moon, Sun, and direction of flight (brown, grey, yellow, and purple vectors).} 
\end{figure}

SPHEREx operates in low-Earth orbit. While this significantly improves the launch cost and access to ground stations when compared with a mission at L2, it does introduce operational complexity that we address with our optimal sensor tasking algorithm. As illustrated in Figure~\ref{fig:avoidance_angles}, the allowable pointing zone is limited by several avoidance angles specified relative both to the observatory and telescope body axes. Note that these constraints vary with time as SPHEREx orbits the earth.

\subsection{Power}

To provide sufficient power to operate the observatory, we developed a requirement on the orbit-averaged solar panel angle. When averaged over the approximately 98-minute orbital period, the solar panel normal vector must be less than $25^\circ$ from the Sun vector. When considering targets for observation in the SPS, we use the history of the solar panel angle over the previous 98 minutes to determine an equivalent (time-varying) instantaneous limit on the solar panel angle to define the allowable pointing zone for the next observation.

\subsection{Thermal}
\label{sec:thermal}

Several avoidance angle requirements in SPHEREx illustrated in Figure~\ref{fig:avoidance_angles} flow in part from thermal considerations. To avoid significantly heating the telescope from illumination by the Sun or Earth, the payload $\hat{z}_{PLD}$ axis needs to be more than 91.5$^\circ$ from the Sun, and also must be within 47.0$^\circ$ of the Earth zenith.

To limit illumination of components mounted outside the Sun shield, we developed a requirement on the overall ``twist'' of the observatory. To define this twist angle, first we consider a fiducial attitude in which the telescope boresight $\hat{z}_{LOS}$ is pointed at the science target, and the observatory is rotated about $\hat{z}_{LOS}$ such that the angle between the solar panel normal $\hat{x}_{PLD}$ and the Sun is minimized. As shown in Figure~\ref{fig:avoidance_angles}, the observatory twist angle is defined as a rotation about $\hat{z}_{LOS}$ relative to this fiducial attitude. Our requirement is to keep this angle below $30^\circ$ at all science attitudes.

\subsection{Telescope Stray Light}

Given the strict thermal requirements of the mission, the Sun and Earth avoidance angle requirements are driven primarily by thermal considerations described in Section~\ref{sec:thermal} and less by telescope stray light considerations. In contrast, the Moon is so bright in the infrared that there is an avoidance angle requirement between the telescope and the Moon based on the stray light performance of the telescope. Simulations show that the direction $\hat{z}_{MOONREF}$ most susceptible to stray light is not perfectly co-aligned with the telescope boresight $\hat{z}_{LOS}$. This means we need to keep the angle between the reference vector $\hat{z}_{MOONREF}$ and the center of the Moon larger than 30$^\circ$ at all science attitudes. Additionally, to prevent direct illumination of the detector arrays that would lead to charge persisting in the detectors onto subsequent exposures, the angle between the telescope boresight $\hat{z}_{LOS}$ and the center of the Moon needs to be larger than 13$^\circ$ at all times, including while slewing the observatory between targets.

\subsection{Shuttle Glow}

During initial operations made with SPHEREx, we observed diffuse shuttle glow \cite{banks83} in the detectors when observing with the telescope pointed close to our direction of flight. We made a series of observations varying the ram angle, that is the angle between the telescope boresight $\hat{z}_{LOS}$ and our direction of flight $\hat{v}$. By measuring the observed diffuse glow as a function of ram angle, we set a limit to keep the ram angle at or above $70^\circ$ at all science attitudes. We selected this angle both to significantly reduce the level of shuttle glow in our images, while leaving a large enough allowable pointing zone to observe efficiently.

\subsection{Pointing Constraint Margin}

The SPS uses an orbit predict to plan survey observations for up to a week in advance. Even using the advanced MONTE\cite{evans18} orbit predict engine at JPL, orbit predict error does accumulate during this period. One impact of this is that the apparent angle of the Sun, Moon, and Earth may be different than intended when at the commanded attitude. To mitigate this orbit predict error, plus any attitude control errors, and to also include margin from onboard fault protection limits, we plan $2^\circ$ inboard of all avoidance angles. This $2^\circ$ value is based on our estimates of the size of all of these error terms, plus margin.

\section{Optimal Target Selection Algorithm}

The core of the survey planning software is our optimal target selection algorithm. Within the finite set of targets that are observable within the allowable pointing zone, in this section we describe the algorithm we use to select the optimal target to observe next.

\subsection{Determination of Allowable Pointing Zone}

As described in Section~\ref{sec:avoidance}, our allowable pointing zone is defined by a number of intersecting, time-varying avoidance angle constraints that each flow directly from power, thermal, and stray light requirements. Due to the number and nature of the constraints, it would be unwieldy to construct the allowable pointing zone directly in an analytic calculation. Instead, before each observing period, and for each target on our main target list, we numerically calculate each avoidance angle assuming SPHEREx was at the attitude required to observe that target. We then select the subset of targets on the list that is within the allowable pointing zone. This typically results in several hundred targets, from which we select the next target to observe with our optimization strategy described in Section~\ref{sec:optimal}.

According to the convention adopted by SPHEREx for the attitude control system, the unrotated orientation of the body axes has the payload $\hat{z}_{PLD}$ axis aligned towards celestial north and the payload $\hat{x}_{PLD}$ aligned towards the RA=0 direction. In this orientation, in J2000 (i.e. Earth-centered equatorial) coordinates the payload body axes are
\begin{eqnarray}
\xpld^\mathrm{init} &=& \hat{x}_\mathrm{ICRF}\nonumber\\
\ypld^\mathrm{init} &=& \hat{y}_\mathrm{ICRF}\nonumber\\
\zpld^\mathrm{init} &=& \hat{z}_\mathrm{ICRF}.
\end{eqnarray}

Aligning the telescope with a desired target requires orienting $\zlos$ to the desired ecliptic longitude and latitude, and orienting $\xlos$ to the desired ecliptic position angle. This is represented by an overall rotation matrix $\mathbf{R_\mathrm{tot}}(\mathrm{lon},\mathrm{lat},\mathrm{PA})$ composed of several intermediate rotations. We derived an analytic expression for this rotation matrix and implemented it in the SPS. This enables us to efficiently precalculate each LOS and PLD body axis for each lon/lat/PA entry on our main target list and store the result in RAM for rapid reuse later in the software.

We then use these body axes to evaluate which targets are within the allowable pointing zone. For example, we evaluate the Sun angle based on the unit vector $\zpld$ corresponding to each target in the main target list. Given the Sun vector $\hat{s}$ at a desired time, we calculate
\begin{equation}
\theta_{\mathrm{Sun}} = \arccos{(\zpld \cdot \hat{s})}
\end{equation}
for the $\zpld$ of each target and only consider targets with $\theta_{\mathrm{Sun}} > 93.5^\circ$ as candidates for observation. We perform similar calculations for all other constraint angles described in Section~\ref{sec:avoidance}.

\subsection{Selecting the Optimal Target for Observation}
\label{sec:optimal}

The main task of the SPS is to select which targets, and in what order, SPHEREx observes during operations. Fundamentally, this sensor tasking problem is an optimization problem. Each observing attitude is defined by three parameters (ecliptic longitude, latitude, and position angle). This means that creating an optimal observing schedule for the hundreds of attitudes in a weeklong observing period is an extremely high-dimensional non-linear optimization problem ($>$1,000 parameters), with complex constraints among all parameters. Since the fully general problem is not tractable due to this combination of high dimensionality and high non-linearity, for SPHEREx we instead formulated the optimization problem as an online problem. In our online approach, at any given moment the SPS only optimizes for the \textit{next} attitude in the survey. With this approximation, the problem therefore reduces to a series of tractable three-dimensional optimizations, and as shown in Section~\ref{sec:coverage} our approach delivers excellent coverage performance for SPHEREx.

\begin{figure}
\begin{center}
\includegraphics[width=0.6\textwidth]{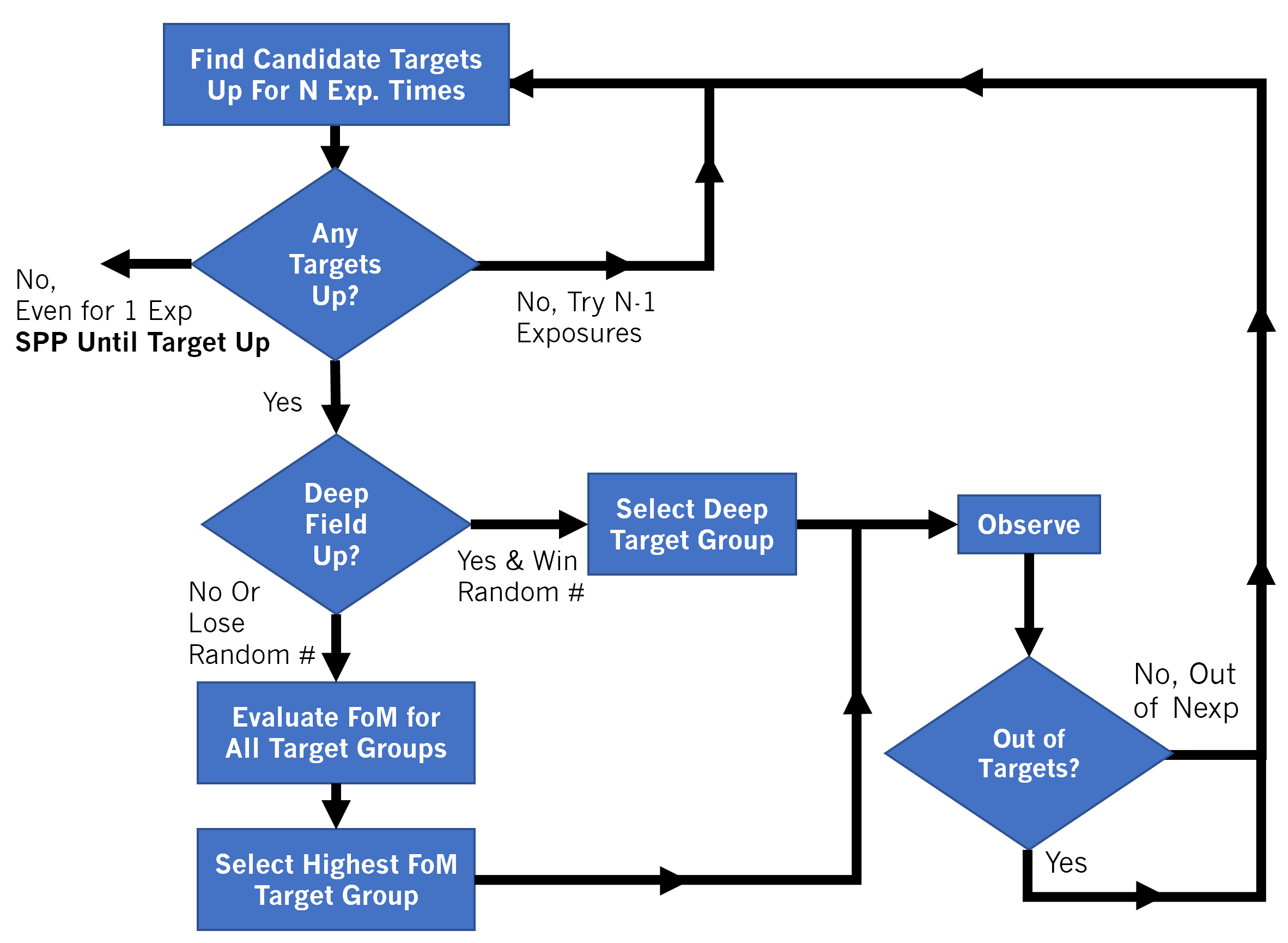} 
\end{center}
\caption 
{ \label{fig:optimal_algorithm}
Flowchart illustrating the SPHEREx SPS target selection algorithm, described in detail in the main text. Starting in the top-left of the flowchart, the algorithm determines which targets are observable. Flowing down to the center-left, the algorithm prioritizes observing the Deep Survey (flowchart branch moving to the center), or observes the All-Sky Survey (flowchart branch moving to bottom left). In either case, after selecting the targets for observation, the algorithm allocates time for observing the selected targets (bottom right of the flowchart) then moves back to the start of the algorithm to select the next targets.} 
\end{figure}

Our online optimization algorithm is illustrated in Figure~\ref{fig:optimal_algorithm}. At a given time $t$ in the mission, the algorithm starts by evaluating what subset of the main target list is currently in the allowable pointing zone. For each large slew we can take up to 4 exposures at 4 different targets within a target group before motion along the orbit results in an Earth constraint violation.  We also evaluate which targets are observable at the end of that time, that is at time
\begin{equation}
t_{\mathrm{later}} = t~+~\mathrm{(Large~Slew~Time)}~+~4\times\mathrm{(Exposure~Time)}~+~3\times\mathrm{(Small~Step~Time)}.
\end{equation}
If no targets are observable at both time $t$ and $t_{\mathrm{later}}$, we try again assuming we make only 3 exposures, then 2, then 1. If no targets are observable even for a single exposure time, we use a different algorithm to calculate a safe place to point (SPP) with a large forward slew, safely passing time while the orbit moves SPHEREx out of these challenging conditions. However, if we find (a) target(s) that can be observed, as illustrated in the flowchart in Figure~\ref{fig:optimal_algorithm} we move forward with the algorithm.

We prioritize observations of the deep fields when they are observable. Thus, if a deep field is observable, the SPS prioritizes the first deep field attitude within the subset of observable targets as the next attitude and jumps to the end of the flowchart to observe the selected target(s). To control the degree to which we prioritize the deep field, the SPS generates a random number and if the number is above a user-defined threshold, the SPS actually selects an All-Sky Survey target instead. Configuring the threshold parameter in the SPS such that it selects the deep field approximately $85\%$ of the times it is observable yields all-sky and deep coverage that meets requirements on both with balanced margin.

If the All-Sky Survey is selected for observation, often dozens to hundreds of targets are observable within the allowable pointing zone. To select which target from this set to observe next, we evaluate a figure of merit (FoM) for each target group and select the optimal target group which has the best FoM. The FoM prioritizes two competing terms, with the relative weighting configurable as a user parameter $F$. On one hand, we prioritize target groups based on the number of observations $N_{obs}$ already made in this target group in earlier orbits. For example, a target group with only 3 out of 17 targets observed would have a better FoM than a target group that already has 16 out of 17 targets observed. On the other hand, we prioritize target groups that are moving in longitude out of the allowable pointing zone since they will not become observable again for months. To calculate this term, driven by the Sun avoidance angle, we first determine which edge of the allowable pointing zone is receding into the Sun, then for each target group we calculate the difference in longitude $\Delta \theta$ between that edge and the target. For example, a target group only 1 degree in longitude away from the receding edge would have a better FoM than a target group 10 degrees from the receding edge. Thus, in total our figure of merit is
\begin{equation}
\mathrm{FoM} = (1-F)\times\frac{17-N_{obs}}{17} + F\times\frac{\Delta \theta}{\theta_{ref}},
\end{equation}
where $\theta_{ref}$ is the average angular size of the entire allowable pointing zone. Based on a series of mission-long simulations, setting the weight parameter to $F\approx0.8$ yields the best survey coverage based on our orbit and typical allowable pointing zone size.

We use a running coverage-to-date file to track $N_{obs}$ for each target group from our target list. The SPS updates this file when observations are planned, and in addition the SPHEREx data analysis pipeline also updates this file if observations are later found to contain excess cosmic ray glitches or other effects that make the observations unusable by the end users. Currently approximately $2\%$ of our observations are found to contain these glitches, and are flagged for reobservation by removing them from our coverage-to-date file. As a natural consequence of the FoM in our algorithm, many of these targets are naturally reobserved. A simulation of this effect is included in the mission-long runs presented in Section~\ref{sec:coverage}.

Having selected which All-Sky Survey or Deep Survey target group to observe, the algorithm allocates the time needed to execute a large slew from the current attitude to the first unobserved target in the target group, then allocates the time needed for the first exposure time and any subsequent small steps to other targets in the target group and subsequent exposures. As shown in Figure~\ref{fig:optimal_algorithm}, the algorithm then returns to the start at the top-left of the flowchart to calculate the next optimal survey attitude.

\section{Achieved Survey Coverage}
\label{sec:coverage}

\begin{figure}
\begin{center}
\includegraphics[width=1.1\textwidth]{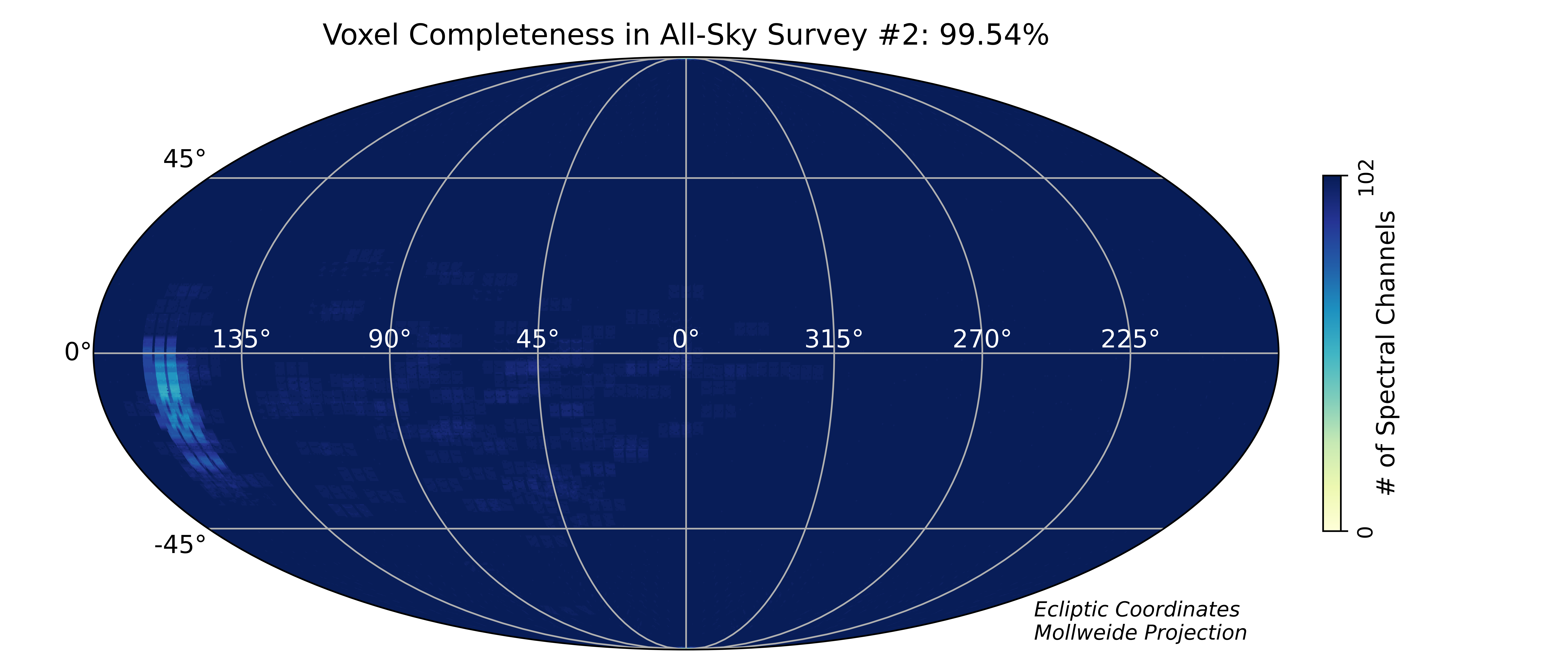} 
\end{center}
\caption 
{ \label{fig:allsky_coverage}
Achieved All-Sky Survey coverage in a simulated 25-month mission planned with the SPHEREx SPS. In this mission-long simulation, each of our 4 All-Sky Surveys achieved 99.54$\%$ coverage (in Survey 2 shown here) or higher (in Surveys 1, 3, and 4 not shown), which represents significant margin against our required coverage of 98$\%$.} 
\end{figure}

On March 12th, 2025, SPHEREx successfully launched into a Sun-synchronous polar orbit from Vandenberg Space Force Base. Our actual trajectory is very similar to the nominal orbit we studied before launch, and science operations are ongoing. Using our actual trajectory, we performed mission-long simulations with the SPS that yield excellent sky coverage. We assess the coverage for the All-Sky Survey using voxel completeness. In the All-Sky Survey, we define a voxel (i.e. spectral datacube volume element) as a single $6.15''\times6.15''$ spatial pixel (out of $41{,}252.96~\mathrm{deg}^2 / (6.15''\times6.15'') \approx 14.14$ billion spatial pixels on the sky) in one spectral channel (out of 102 in our sensor system). This means the total number of voxels in a SPHEREx All-Sky Survey is (14.14 billion spatial pixels$)\times$(102 spectral channels) $\approx$ 1.442 trillion voxels.

We developed assessment software that reads in a survey plan generated with the SPS for our entire mission, calculates the number of voxels observed by that survey plan, and divides by the total number of voxels on the sky to yield a metric we define as voxel completeness. Many effects will limit the end-to-end voxel completeness of our mission, including a small number of defective pixels in our detector arrays, rare data loss due to errors in data downlink, lost observing time due to spacecraft anomalies, and performance of the SPS. As part of an end-to-end budget of the mission voxel completeness, the SPS is required to deliver survey plans with at least 98$\%$ planned voxel completeness. In this paper, we present the planned coverage delivered by the SPS, and track other losses to voxel completeness elsewhere. As shown in Figure~\ref{fig:allsky_coverage}, with a nominal orbit predict the SPS delivers planned voxel completeness of $99.54\%$, representing significant margin on our $98\%$ requirement.

\begin{figure}
\begin{center}
\includegraphics[width=1.0\textwidth]{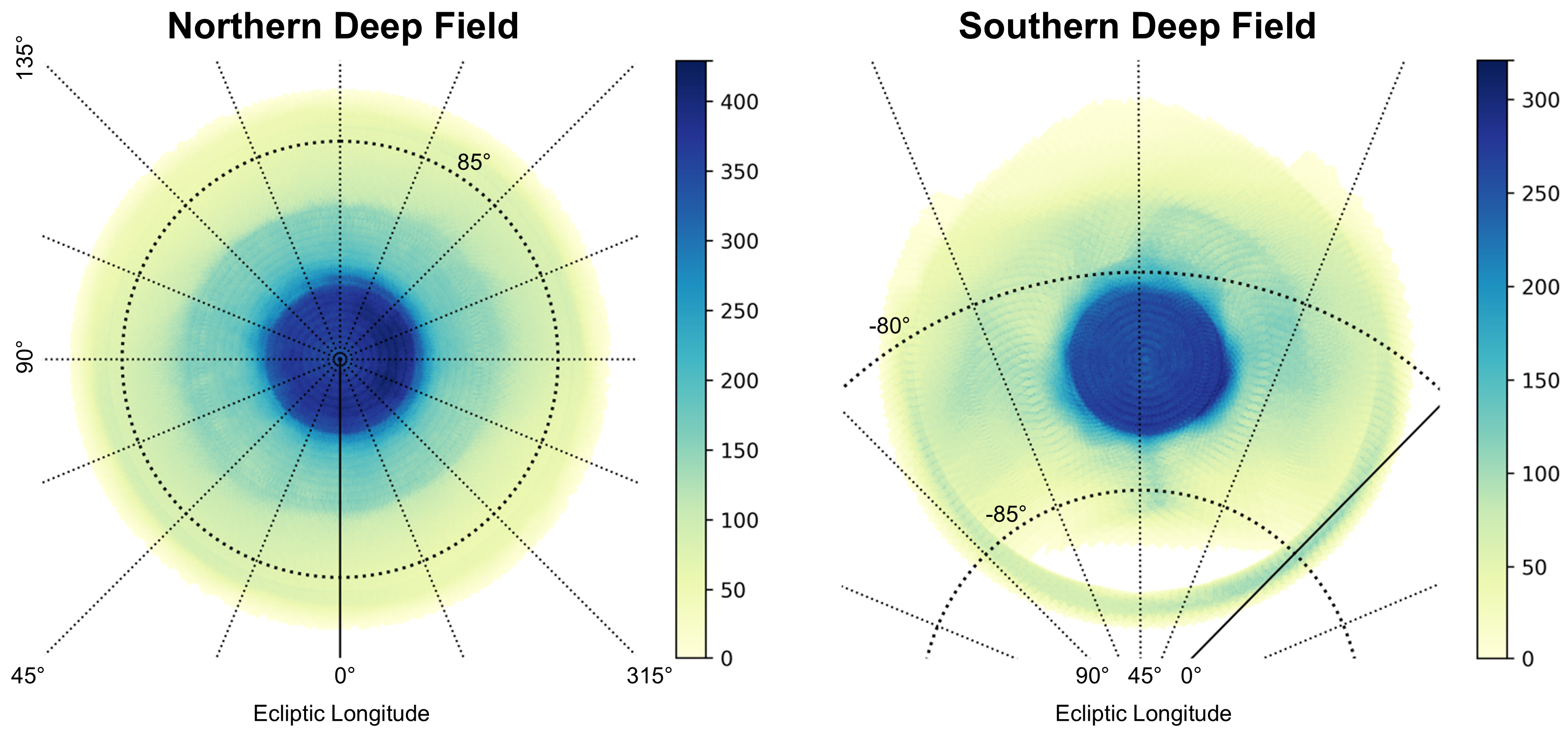} 
\end{center}
\caption 
{ \label{fig:deep_coverage}
Achieved coverage in the northern (left) and southern (right) fields of the Deep Survey in a simulated 25-month mission planned with the SPHEREx SPS. The color shows the number of complete spectra per 6.15''$\times$6.15'' pixel on the sky. When assessed against the deep field coverage metric in the mission requirements, the combined north+south coverage meets the requirements with significant margin. Going above the requirements, our coverage in the northern deep field alone also meets the coverage requirements.} 
\end{figure}

As shown in Figure~\ref{fig:deep_coverage}, we also achieve excellent coverage in the deep fields. The coverage maps show we will have hundreds of fully-sampled independent spectra for each $6.15''\times6.15''$ spatial pixel in the central 3.5-degree diameter core area of each deep field. To quantitatively assess our coverage against the SPHEREx science requirements\cite{dore16}, we developed a metric to assess the uncertainty $\sigma\sqrt{D_l}$ of our measurement of the angular power spectrum of EBL made using our deep field maps. Using the Knox power spectrum sensitivity forecast formalism\cite{knox95}, the $i$-th sky area element with solid angle $a_i$ contributes a variance (relative to the detector noise level $s$) of
\begin{equation}
\sigma^2 D_{l,i} / s^2 = \frac{\Omega}{N_{hit,i} N_{ch}} \frac{l (l+1)}{2 \pi} \sqrt{\frac{2}{2l + 1}\frac{4 \pi}{a_i}},
\end{equation}
to the power spectrum. Here $l$ is the multipole moment on the sky, $N_{hit,i}$ is the number of observations in the $i$-th sky area element (i.e. the coverage maps in Figure~\ref{fig:deep_coverage}), $N_{ch}=8$ is the number of spectral channels we average together to yield the broader-bandwidth EBL spectral channels referenced in our requirements, and $\Omega$ is the solid angle on the sky of a single pixel. We integrate this first against the area elements from the coverage maps in Figure~\ref{fig:avoidance_angles}
\begin{equation}
K_l = \sum_i \left( \sigma^2 D_{l,i} \right)^{-2}
\end{equation}
then over the 5-20 arcminute angular scales (i.e. 500-2000 in multipole $l$) of interest
\begin{equation}
\sigma^2 D_l = \Delta l \sum_{l=500}^{2000} \left( K_l \right)^{-1/2}
\end{equation}
with $\Delta l = 100$ based on the spatial extent of our coverage region, to yield our final deep-field sensitivity metric $\sigma\sqrt{D_l} / s$. Based on our science goals\cite{dore16}, we have a requirement to deliver an uncertainty level of $40\times10^{-6}$ or lower in these units. In our mission-long simulation, we achieve a sensitivity of $29.3\times10^{-6}$ when combining data from both deep fields, which represents significant margin against our requirement. In addition our northern deep field would meet our full-mission requirement alone, since it achieves a sensitivity metric of $37.9\times10^{-6}$.

\section{Conclusions}

The SPHEREx Survey Planning Software is one of many new capabilities we developed for our mission that enable us to deliver transformational science within the challenging operational constraints of low-Earth orbit. Our algorithm selects targets for efficient observation to optimize the total coverage we achieve during mission operations. The resulting spectral coverage meets our mission requirements with significant margin. As demonstrated by our successful first public data release\cite{spherex25}, the survey planning software is performing well in flight.

\subsection* {Acknowledgments}
We acknowledge support from the SPHEREx project under a contract from the NASA/Goddard Space Flight Center to the California Institute of Technology. Part of the research described in this paper was carried out at the Jet Propulsion Laboratory, California Institute of Technology, under a contract with the National Aeronautics and Space Administration (80NM0018D0004).

%%%%% References %%%%%

\bibliography{article}   % bibliography data in report.bib
\bibliographystyle{spiejour}   % makes bibtex use spiejour.bst

\end{document}